\documentclass[aps,amssymb,floatfix,prd,amsmath,preprintnumbers]{revtex4-2}
\usepackage{float}
\usepackage{capt-of}
\usepackage[svgnames]{xcolor}
\usepackage{hyperref}
\hypersetup{
  colorlinks = true,
  linkcolor =Red,
  anchorcolor = red,
  citecolor = red,
  urlcolor = yellow
}
\usepackage{graphicx}  % Include figure files
\usepackage{dcolumn}   % Align table columns on decimal point
\usepackage{bm}% bold math
\begin{document}
%\input epsf.tex
%%%%%%%%%%%%
%%%%%%%%%%%

\title{\bf Model parameters in the context of late time cosmic acceleration in $f(Q,T)$ gravity}

\author{Laxmipriya Pati\footnote{Department of Mathematics, Birla Institute of Technology and Science-Pilani, Hyderabad Campus, Hyderabad-500078, India, E-mail:lpriyapati1995@gmail.com}, B. Mishra\footnote{Department of Mathematics, Birla Institute of Technology and Science-Pilani, Hyderabad Campus, Hyderabad-500078, India, E-mail:bivu@hyderabad.bits-pilani.ac.in}, S.K. Tripathy\footnote{Department of Physics, Indira Gandhi Institute of Technology, Sarang, Dhenkanal, Odisha-759146, India, E-mail:tripathy\_sunil@rediffmail.com}
}

\affiliation{ }

\begin{abstract}
The dynamical aspects of some accelerating models are investigated in the framework of an extension of symmetric teleparllel gravity dubbed as $f(Q,T)$ gravity. In this gravity theory, the usual Ricci tensor in the geometrical action is replaced by a functional $f(Q,T)$ where $Q$ is the non-metricity and $T$ is the trace of the energy-momentum tensor. Two different functional forms are considered in the present work. In order to model the Universe, we have considered a signature flipping deceleration parameter simulated by a hybrid scale factor (HSF). The dynamical parameters of the model are derived and analysed. We discuss the role of the parameter space in getting viable cosmological models. It is found that, the models may be useful as suitable geometrical alternatives to the usual dark energy approach.
\end{abstract}

\maketitle

\section{Introduction} 

At the classical level, several approaches are presented to address the observational results concerning the structure, formation and dynamics of the Universe, but the search for a satisfactory gravity theory is still to be established. In this context, a recently proposed gravity theory is the $f(Q,T)$ gravity, where $Q$ and $T$ respectively represent the non-metricity and the trace of energy momentum tensor $T_{\mu\nu}$ \cite{Xu19}. This is an extension of symmetric teleparallel gravity. 
The basic idea in teleparallel gravity is to replace the basic physical variable, the metric $g_{\mu\nu}$ in the space-time, by the set of tetrad vectors $e^i_{\mu}$ and in turn, the tetrad vectors generate the torsion. Now, the curvature has been replaced by the torsion, which will provide the gravitational effects of the Universe. This approach is known as the teleparellel equivalent of GR and most commonly known as $f(T)$ gravity \cite{Moller61,Pellegrini63,Havashi79}. Aldrovandi and Pereira have presented a systematic discussions on the teleprallel gravity \cite{Aldrovandi13}. Several cosmological models pertaining to the issues of cosmology and astrophysics have been obtained recently. Ferraro and Fiorini \cite{Ferraro07} have solved the particle horizon problem in flat FRW space-time. Linder \cite{Linder10} has shown the result for cosmic acceleration with the behaviour of the effective equation of state parameter with respect to the function, which describes the exponential dependency on torsion scalar. Harko et al. \cite{Harko14} have obtained the initial inflationary phase, the matter-dominated expansion, and then the late-time accelerating phase in $f(T,\mathcal{T})$ gravity. Capozziello et al. \cite{Capozziello17} have used the observational data of Big Bang Nucleosynthesis (BBN) to constrain the $f(T)$ gravity model for the power law, exponential law and square root exponential law. Otalora and Reboucas \cite{Otalora17} have presented the Godel type geometries and its solution in $f(T)$ gravity. Golovnev and Koivisto \cite{Golovnev18} have performed the cosmological perturbation analysis in modified $f(T)$ gravity. Fontanini et al. \cite{Fontanini19} have critically commented on the teleparallel equivalent of GR. Bose and Chakraborty \cite{Bose20} have studied the cosmological aspects and the thermodynamical analysis of the model framed with $f(T)$ gravity. Golovnev and Guzman have \cite{Golovnev20} have shown the Bianchi identities for $f(T)$ gravity and applied to obtain the spherically symmetric solutions.\\

According to the literature available, GR can be represented  in teleparallel representation and curvature representation. Both the geometric representation are equivalent; in teleparallel representation, the curvature and non-metricity vanishes whereas in curvature representation the torsion and non-metricity vanishes. Another equivalent approach is the symmetric teleparallel gravity (STG), originally introduced by Nester and Yo\cite{Nester99}, where the basic geometry of the gravitational action is represented by the non-metricity $Q$ of the metric. Jimenez et al. \cite{Jimenez18}  further developed the STG to the coincident GR or the $f(Q)$ gravity. This is also known as the non-metric gravity. In STG, Soudi et al. \cite{Soudi18} have  indicated that the gravitational wave polarizations are instrumental in constraining the strong field behaviour of the theory of gravity. Using red shift function, Lazkoz et al. \cite{Lazkoz19} have analysed the observational results in $f(Q)$ gravity models. Lu et al. \cite{Lu19} have derived the field equations of $f(Q)$ gravity and  have shown that the equation of state parameter can cross over the phantom divide line. Bajardi et al. \cite{Bajardi20} have shown the bouncing behaviour of the cosmological model in $f(Q)$ gravity. Frusciante \cite{Frusciante21} has  investigated the role of $f(Q)$ gravity in the context of cosmological observables. The $f(Q,T)$ gravity is another extension of $f(Q)$ gravity where the Lagrangian density of gravitational field is described through, $L=f(Q,T)$ \cite{Xu19}. In the $f(Q,T)$ gravity, a non-minimal coupling between the non-metricity $Q$ and the trace $T$ of the energy-momentum tensor is assumed. The motivation behind such a geometrically extended gravity is to address the late time cosmic phenomena issue. Xu et al. \cite{Xu20} also suggested the Weyl type $f(Q,T)$ gravity and its cosmological implications. Zia et al. \cite{Zia21} have obtained the cosmological model in $f(Q,T)$ gravity applying the energy conservation equation.\\ 

The paper is organised as follows: in Sec II, the action and basic field equations of $f(Q,T)$ gravity have been presented and the dynamical parameters are derived. In Sec III, we have investigated the cosmological formalism of two models with the presumed hybrid scale factor (HSF). The dynamical behaviour of both the models along with the graphical representation of the dynamical parameters are presented. Results and conclusions of the models are given in Sec IV. 

\section{Field Equations in $f(Q,T)$ gravity}
The action for the extended symmetric teleparallel gravity, the $f(Q,T)$ gravity is given by \cite{Xu19},
\begin{equation} \label{eq.1}
S=\int\left[\dfrac{1}{16\pi}f(Q,T)+\mathcal{L}_{m}\right]d^{4}x\sqrt{-g},
\end{equation}
where $f(Q,T)$ represents the arbitrary function of the non-metricity $Q$ and trace of the energy momentum tensor $T$. The matter Lagrangian is represented by $\mathcal{L}_m$ and $g=det(g_{ij})$ be the determinant of the metric tensor $g_{ij}$. The non-metricity may be defined as,
\begin{equation}
Q\equiv -g^{\mu \nu}( L^k_{~l\mu}L^l_{~\nu k}-L^k_{~lk}L^l_{~\mu \nu}),
\end{equation}
where $L^k_{~l\gamma}\equiv-\frac{1}{2}g^{k\lambda}(\bigtriangledown_{\gamma}g_{l\lambda}+\bigtriangledown_{l}g_{\lambda \gamma}-\bigtriangledown_{\lambda}g_{l\gamma})$.\\

By varying the gravitational action \eqref{eq.1}, the field equation of $f(Q,T)$ gravity \cite{Xu19} can be obtained as, 
\begin{equation}\label{eq.2}
-\frac{2}{\sqrt{-g}}\bigtriangledown_{k}(f_{Q}\sqrt{-g}p^{k}_ {\mu \nu})-\frac{1}{2}fg_{\mu \nu}+f_{T}(T_{\mu \nu}+\Theta_{\mu \nu})-f_{Q}(p_{\mu kl} Q^{\;\;\; kl}_{\nu}-2Q^{kl}_{\;\;\;\mu} p_{kl\nu})=8 \pi T_{\mu \nu},
\end{equation}
where $f_Q=\frac{\partial f}{\partial Q}$ and we represent $f\equiv f(Q,T)$. The super potential of the model is defined as, $p^{k}_{\mu \nu}=-\frac{1}{2}L^{k}_{\mu \nu}+\frac{1}{4}(Q^{k}-\tilde{Q}^{k})g_{\mu \mu}-\frac{1}{4}\delta^{k}_{(\mu}Q_{\nu)}$; and the energy momentum tensor, $T_{\mu \nu}=\frac{-2}{\sqrt{-g}} \frac{\delta(\sqrt{-g}L_{m})}{\delta g^{\mu \nu}}$ and $\Theta_{\mu \nu}=g^{kl}\frac{\delta T_{kl}}{\delta g^{\mu \nu}}$. $T=T_{\mu \nu}g^{\mu \nu}$ is the trace of the energy momentum tensor and that of the non-metricity is, $Q_{k}=Q_{k}^{\;\;\mu}\;_{\mu}$, $\tilde{Q}_{k}=Q^{\mu}\;_{k\mu}$. We have considered the homogeneous, isotropic and spatially flat FLRW metric in the form, 

\begin{eqnarray}\label{eq.3}
ds^{2}=-N^{2}(t)dt^{2}+\mathcal{R}^{2}(t)(dx^{2}+dy^{2}+dz^{2}),
\end{eqnarray}
where $N(t)$ and $\mathcal{R}(t)$ are respectively the lapse function and scale factor. The Hubble function that describes the expansion rate can be related to the scale factor as $H(t)=\frac{\dot{\mathcal{R}}(t)}{\mathcal{R}(t)}$, where an over dot is the derivative with respect to cosmic time $t$. Also, the dilation rate can be defined as, $\tilde{T}=\frac{\dot{N}(t)}{N(t)}$. It can be obtained in a straightforward manner that, $Q=6\frac{H^2}{N^2}$. For the standard FRW case, the lapse function is assumed to be $N(t)=1$ so that the non-metricity becomes $Q=6H^2$. Also for the standard case, the dilation rate vanishes. We consider a perfect fluid distribution in the Universe so that, the energy momentum tensor is expressed as, $T_{\mu \nu}=diag(-\rho,p,p,p)$. Subsequently, we get, $\Theta_{\mu \nu}=diag(2\rho+p,-p,-p,-p)$. 

Now, the field equations \eqref{eq.2} of $f(Q,T)$ gravity in the standard case ($N=1$) for the FLRW space-time can be obtained as,

\begin{eqnarray}
p&=&-\frac{1}{16\pi}\left[f-12FH^2-4\dot{\chi}\right],\label{4} \\
\rho&=&\frac{1}{16\pi}\left[f-12F H^2-4\dot{\chi}\kappa_1\right],\label{5}
\end{eqnarray}
where $F\equiv \frac{\partial f}{\partial Q}$, $8\pi\kappa\equiv f_T=\left(\frac{\partial f}{\partial T}\right)$, $\kappa_1=\frac{\kappa}{1+\kappa}$ and $\chi=FH$.
We can retrieve the evolution equation of the Hubble function by adding eqns. \eqref{4} and \eqref{5} as,

\begin{equation} \label{eq.6}
\dot{\chi}-4\pi\left(\rho+p\right)\left(1+\kappa\right)=0.
\end{equation}

Now, the equivalent Friedman equations for the present gravity theory may be written as, 
\begin{eqnarray}
2\dot{H}+3H^2&=&\frac{1}{F}\left[\frac{f}{4}-2\dot{F}H+4\pi\left[(1+\kappa)\rho+(2+\kappa)p\right]\right]=-8\pi p_{eff},\label{eq.7} \\
3H^2&=& \frac{1}{F}\left[\frac{f}{4}-4\pi\left[(1+\kappa)\rho+\kappa p\right]\right]=8\pi \rho_{eff}.  \label{eq.8}
\end{eqnarray}
Obviously, the effective energy density $\rho_{eff}$ and the effective pressure $p_{eff}$ satisfy the conservation equation
\begin{equation}
\dot{\rho}_{eff}+3H\left(\rho_{eff}+p_{eff}\right)=0.
\end{equation}

Since the deceleration parameter $q$ is defined as $q=-1-\frac{\dot{H}}{H^2}$, from the above equations, it may be obtained in a straightforward manner that
\begin{equation}
q=-1+\frac{3\left(4\dot{F}H-f+16\pi p\right)}{f-16\pi\left[\left(1+\kappa\right)\rho+\kappa p\right]}.
\end{equation}

The equation of state (EoS) parameter  $\omega=\frac{p}{\rho}$ for the new gravity theory may be obtained as
\begin{equation}
\omega=-1+\frac{4\dot{\chi}}{\left(1+\kappa\right)\left(f-12F H^2\right)-4\dot{\chi}\kappa}.
\end{equation}

In order to investigate viable cosmological scenario in the framework of the above discussed $f(Q,T)$ gravity theory, it is required to consider certain assumed form of the functional $f(Q,T)$. In the seminal work, Xu et al. \cite{Xu19} have considered three different forms for $f(Q,T)$ such as (i) $f(Q,T)= aQ+bT$  (ii) $f(Q,T)=aQ^{m}+bT$ and (iii) $f(Q,T)=-\left(aQ+bT^2\right)$. Here $a,b$ and $m$ are constants. One should note that, the $f(Q,T)$ gravity theory with these functional forms reduce to the usual $f(Q)$ and $f(T)$ theory with the substitution $a=0$ and $b=0$ respectively. In the next section, we will consider the first two cases where the functional contains a minimal coupling of the non-metricity $Q$ and the trace $T$ of the energy momentum tensor. Also, the functional $f(Q,T)$ contains a term linear in $T$.
\section{Cosmological Models}
For a given functional form of $f(Q,T)$, it is required to solve the system of eqns. \eqref{eq.7} and \eqref{eq.8} to get viable cosmological models. However obtaining an exact solution becomes a cumbersome process because of the non linearity nature of the equivalent Friedman equations.  In view of this, either we need to assume a specific equation of state (assumed relationship between the pressure and energy density) to obtain the scale factor or we may consider an assumed dynamics to obtain the dynamically changing equation of state parameter. Here, we wish to go with the second option and consider a specific form of the Hubble parameter as $H=\alpha+\frac{\beta}{t}$ and then investigate the dynamical aspects of the model.  The Hubble parameter provides us a hybrid scale factor(HSF) $\mathcal{R}(t)=e^{\alpha t}t^{\beta}$ and a signature flipping deceleration parameter as, $q=-1+\frac{\beta}{(\alpha t+\beta)^2}$. This assumption provides $q \simeq -1+\frac{1}{\beta}$ at $t\rightarrow 0$ and $q\simeq-1$ at $t\rightarrow \infty$.  The cosmological aspect of the HSF has already been investigated in some earlier works \cite{Mishra15,Mishra18a,Mishra19,Mishra19a,Ray19,Mishra21,Tripathy2020,SKT2020,Mishra18}. It is worth to mention here that, the HSF provides us a deceleration parameter which assumes early positive values and late epoch negative values. In fact, such a signature flipping behaviour of the deceleration parameter is required to mimic the present Universe with late time cosmic acceleration. Also, after the advent of the $H_0$ tension, it is believed that, a transitive deceleration parameter fostering an early deceleration and late time acceleration is in accordance with the concordance $\Lambda$CDM model \cite{Reiss2016,Reiss2018,Aghanim18,Reid2019}. For a brief review of the cosmological models with HSF, one may see Ref. \cite{SKT2021}. Recently, Tripathy et al. \cite{SKT2020, Tripathy2020}  have constructed four HSF models basing upon the $H(z)$ data. These models are termed as HSF11, HSF21, HSF12 and HSF22 respectively. While the HSF11 and HSF21 models predict the transition redshift  to be $z_{da}=0.8$, the other two models predict $z_{da}=0.5$. These values of cosmic transit redshift are in conformity with recent observations \cite{Busca13, Farooq13, Moresco16}. In the present work, we will consider these four HSF models within the framework of $f(Q,T)$ gravity to investigate the dynamical aspects. In order to have an idea about the HSF models, we have presented the deceleration parameter as a function of redshift $z$ in FIG. 1. We have defined the redshift as $z=-1-\frac{\mathcal{R}_0}{\mathcal{R}}$, where $\mathcal{R}_0$ is the scale factor of the Universe at the present epoch. The HSF models predict the deceleration parameter at the present epoch as $q_0=-0.617, -0.723, -0.494$ and $-0.603$ respectively.

\begin{figure}[!htp]
\centering
\includegraphics[scale=0.40]{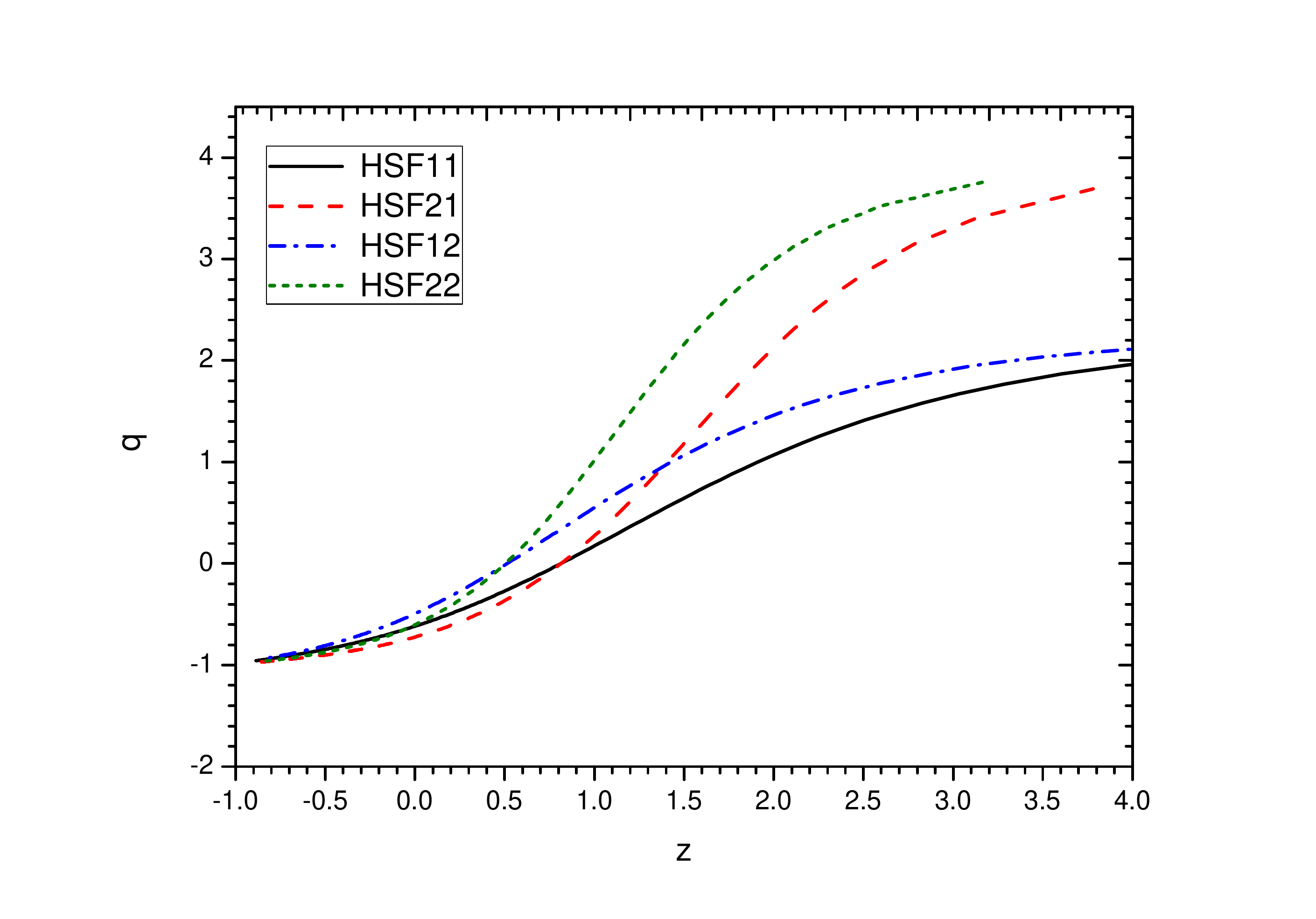}
\caption{Deceleration parameter for the HSF models.}
\label{Fig1}
\end{figure}
Using the HSF, we can obtain the pressure and energy density from the field equations \eqref{4} and \eqref{5} as,

\begin{eqnarray}
p&=&-\frac{1}{16\pi}\left[f-12F \left(\alpha+\frac{\beta}{t}\right)^2-4\dot{F}\left(\alpha+\frac{\beta}{t}\right)+4F\left(\frac{\beta}{t^2}\right)\right]\label{eq.9}, \\
\rho&=&\frac{1}{16\pi}\left[f-12F \left(\alpha+\frac{\beta}{t}\right)^2-\left\lbrace \dot{F}\left(\alpha+\frac{\beta}{t}\right)-F\left(\frac{\beta}{t^2}\right)\right\rbrace4\kappa_1\right].\label{eq.10}
\end{eqnarray}

In the following, we shall consider two different functional forms of $f(Q,T)$ to obtain the dynamical parameters of the cosmological model.  
\subsection{Case I}
We consider the linear form of the functional in the form $f(Q,T)=aQ+bT$ such that $F=\frac{\partial f}{\partial Q}=a$ and $\kappa=\frac{b}{8\pi}$. So the pressure and energy density can be obtained as,

\begin{eqnarray}
p &=& \frac{3aH^2+a\dot{H}\left[2+\kappa-\kappa \kappa_1\right]}{2\pi\left[\left(2+\kappa\right)^2+4\kappa-\kappa^2\right]},\\
\rho &=& \frac{-3aH^2+a\dot{H}\left[3\kappa-\left(2+3\kappa\right)\kappa_1\right]}{2\pi\left[\left(2+\kappa\right)^2+4\kappa-\kappa^2\right]}.
\end{eqnarray}
%\begin{eqnarray}
%p&=&\dfrac{-12a\left(\alpha+\dfrac{\beta}{t}\right)^2(b+8 \pi)+4a(3b+16 \pi)\left(\frac{\beta}{t^2}\right)}{b^2-(3b+16 \pi)^2},\label{eq.11}\\ 
%\rho&=&\dfrac{12a\left(\alpha+\dfrac{\beta}{t}\right)^2 (b+8 \pi)+4ab\left(\frac{\beta}{t^2}\right)}{b^2-(3b+16 \pi)^2}. \label{eq.12}
%\end{eqnarray}
% We may note here that the NEC $p+\rho=\frac{16a(b+4\pi)\frac{\beta}{t^2}}{b^2-(3b+16\pi)^2}$  can be violated when $b^2<(3b+16\pi)^2$. 
For the HSF, we have $H=\alpha+\frac{\beta}{t}$ and $\dot{H}=-\frac{\beta}{t^2}$. Consequently, the pressure and energy density for the present model may be expressed as
\begin{eqnarray}
p &=& \frac{3a\left(\alpha+\frac{\beta}{t}\right)^2-a\left(2+\kappa-\kappa \kappa_1\right)\frac{\beta}{t^2}}{2\pi\left[\left(2+\kappa\right)^2+4\kappa-\kappa^2\right]},\\
\rho &=& \frac{-3a\left(\alpha+\frac{\beta}{t}\right)^2-a\left[3\kappa-\left(2+3\kappa\right)\kappa_1\right]\frac{\beta}{t^2}}{2\pi\left[\left(2+\kappa\right)^2+4\kappa-\kappa^2\right]}.
\end{eqnarray}

\begin{figure}[!htp]
\centering
\includegraphics[scale=0.40]{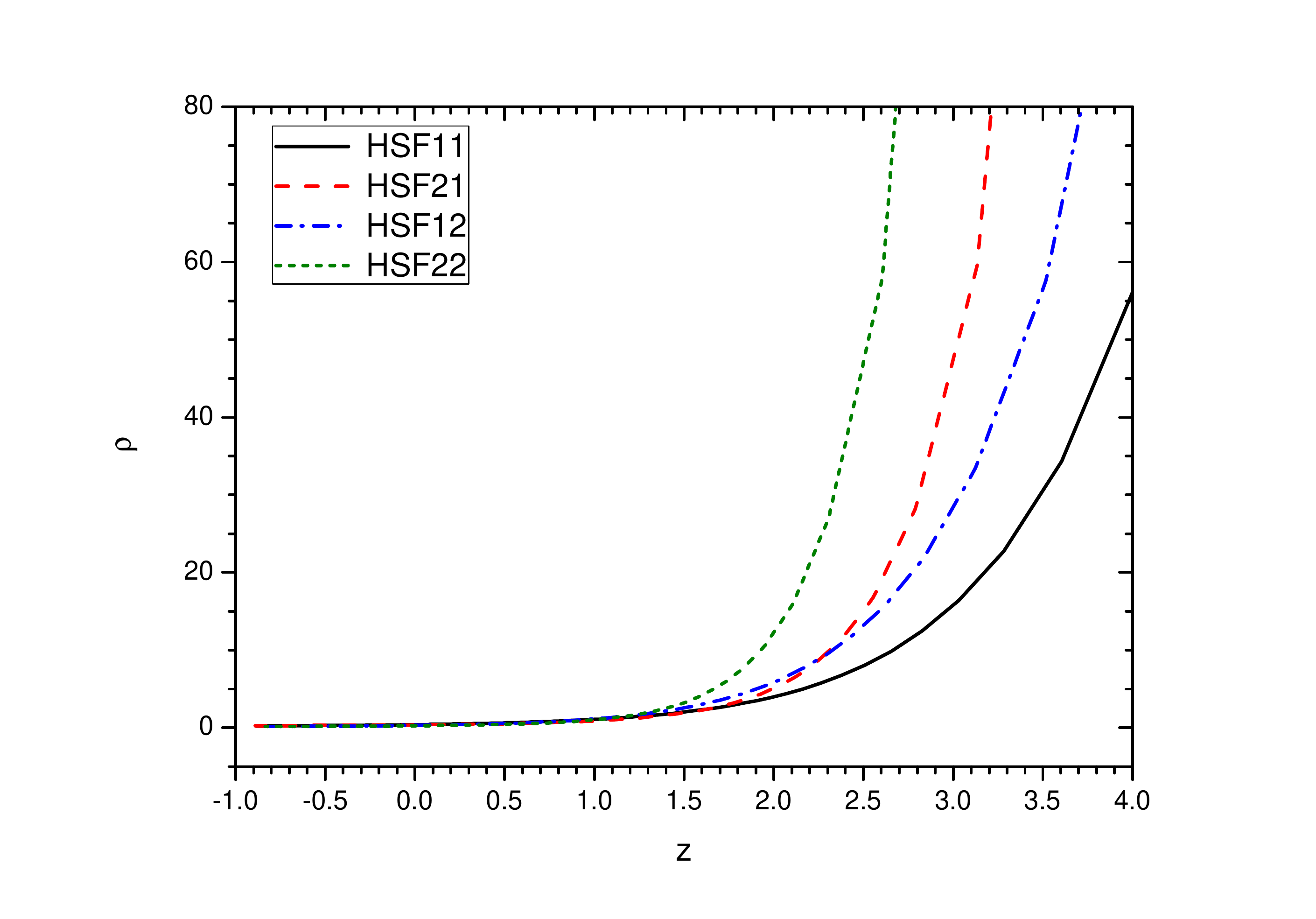}
\caption{ The evolutionary aspect of the energy density for the HSF models. Here we have used the parameter space $a=-4.4, b=0.01$.}
\label{Fig2}
\end{figure}

The evolutionary aspects of the pressure and the energy density obviously depend on the choice of the model parameters $\alpha, \beta, a$ and $b$. The HSF parameters $\alpha$ and $\beta$ are taken from some earlier works \cite{Tripathy2020, SKT2020}. The $f(Q,T)$ gravity parameters $a$ and $b$ are fixed from the behaviour of the energy density and the equation of state $p=p(\rho)$. In order to ensure positivity of the energy density we consider a negative value for $a$ i.e. $a=-4.4$. The other parameter $b$ is considered as $b=0.01$ to get a smooth behaviour of the pressure. The graphical behaviour of the energy density is shown in FIG. 2. For all the HSF models, the energy density becomes positive and smoothly decreases from high positive values to vanishingly small values at late times. The rate of decrement in the energy density defines the characteristics of the HSF models.

Once the pressure and energy density are obtained, it becomes straightforward to calculate the EoS parameter $\omega=\frac{p}{\rho}$ as,
\begin{equation}
\omega=-\frac{3\left(\alpha+\frac{\beta}{t}\right)^2-\left(2+\kappa-\kappa \kappa_1\right)\frac{\beta}{t^2}}{3\left(\alpha+\frac{\beta}{t}\right)^2+\left[3\kappa-\left(2+3\kappa\right)\kappa_1\right]\frac{\beta}{t^2}}.\label{eq.13}
\end{equation}

Several cosmological observations have constrained the numerical value of the EoS parameter such as: Supernovae Cosmology Project, $\omega=-1.035^{+0.055}_{-0.059}$ \cite{Amanullah10} ; WMAP+CMB, $\omega=-1.073^{+0.090}_{-0.089}$ \cite{Hinshaw13}; Planck 2018, $\omega=-1.03\pm 0.03$ \cite{Aghanim18}. We have presented below the graphical behaviour of the EoS parameter $\omega$ in FIG.3, by appropriately choosing the model parameters in order to make a comparison with the already obtained observational results. The EoS parameter evolves from early positive values to almost overlap with the concordant $\Lambda$CDM  value $\omega=-1$ at late times. The rate of evolution for the EoS parameter is decided by the $\beta$ value of the HSF model.  One may observe that, higher the value of $\beta$, lower is rate of decrement of the EoS parameter. It is interesting to note that, the EoS parameter does not depend on the choice of the $f(Q,T)$ parameter $a$. However, it does depend on the other $f(Q,T)$ parameter $b$ but marginally. We have investigated the role of $b$ in the dynamics of $\omega$ and found that, for low values of $b$ in the range $b<10$, there is almost no change of the present epoch EoS parameter value of $\omega=-0.745$ for the model HSF11. On the other hand, if we increase the value of $b$ beyond 10, $\omega$ is found to increase slowly to $-0.33$. Basing upon the behaviour of the EoS parameter, we may conclude that, the present HSF models within $f(Q,T)$ gravity favours a quintessence like evolutionary phase.

\begin{figure}[!htp]
\centering
\includegraphics[scale=0.40]{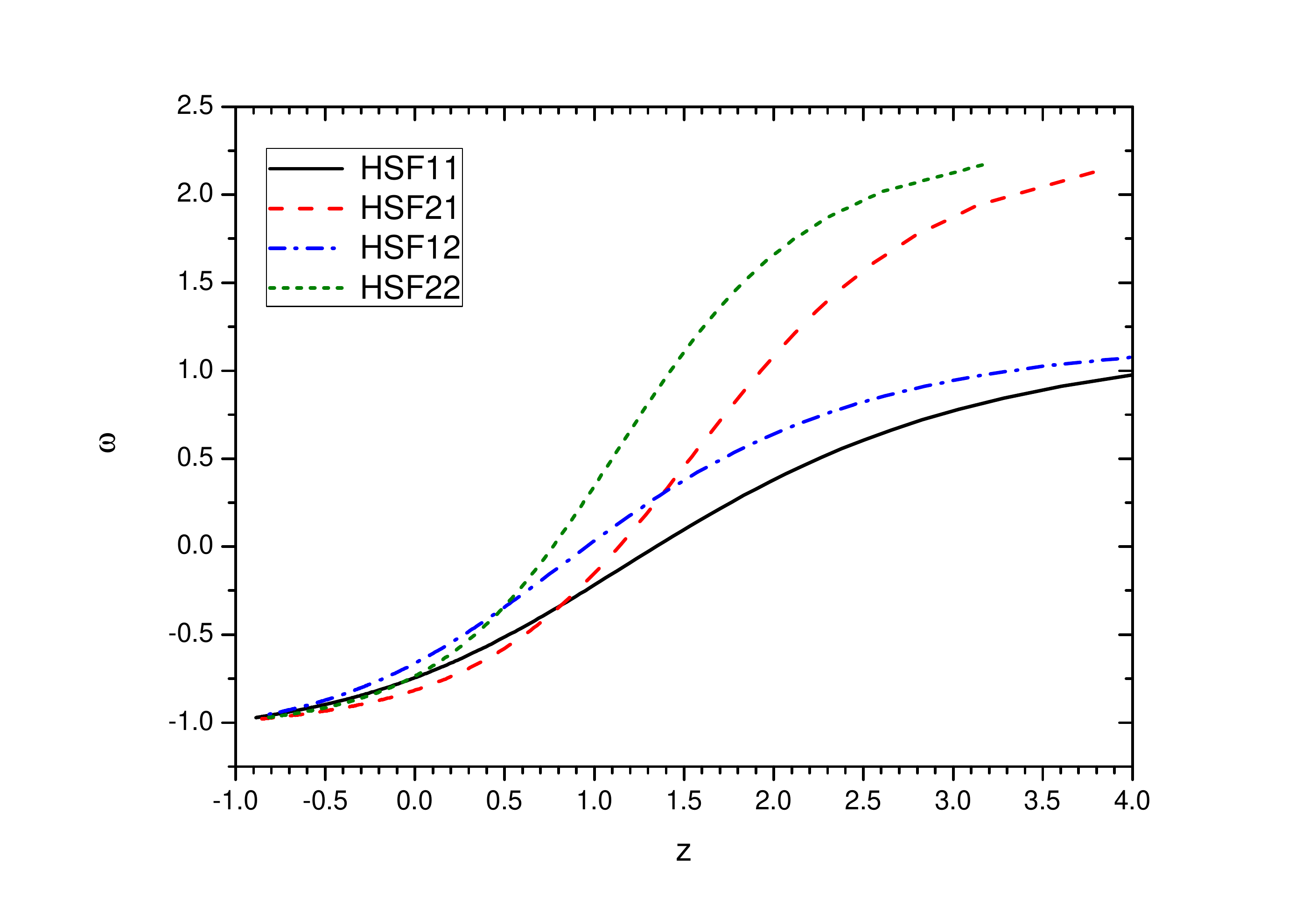}
\caption{The equation of state parameter in $f(Q,T)$ gravity. The parameter space used for the figure is $\left(a=-4.4, b=0.01\right)$.}
\label{Fig3}
\end{figure}

In order to understand the behaviour of the functional $f(Q,T)$ in the dynamical behaviour of the models, we have presented the plot of $f(Q,T)$ for the four HSF models in FIG. 4. One may note that, all the HSF model predict almost same values for the functional $f(Q,T)$ at late times of cosmic evolution. However, at an initial epoch, each HSF model predicts a different evolutionary picture for $f(Q,T)$. One may note that, models with lower values of $\beta$ provide stiff decrement in  the functional $f(Q,T)$ at an early epoch.
\begin{figure}[!htp]
\centering
\includegraphics[scale=0.40]{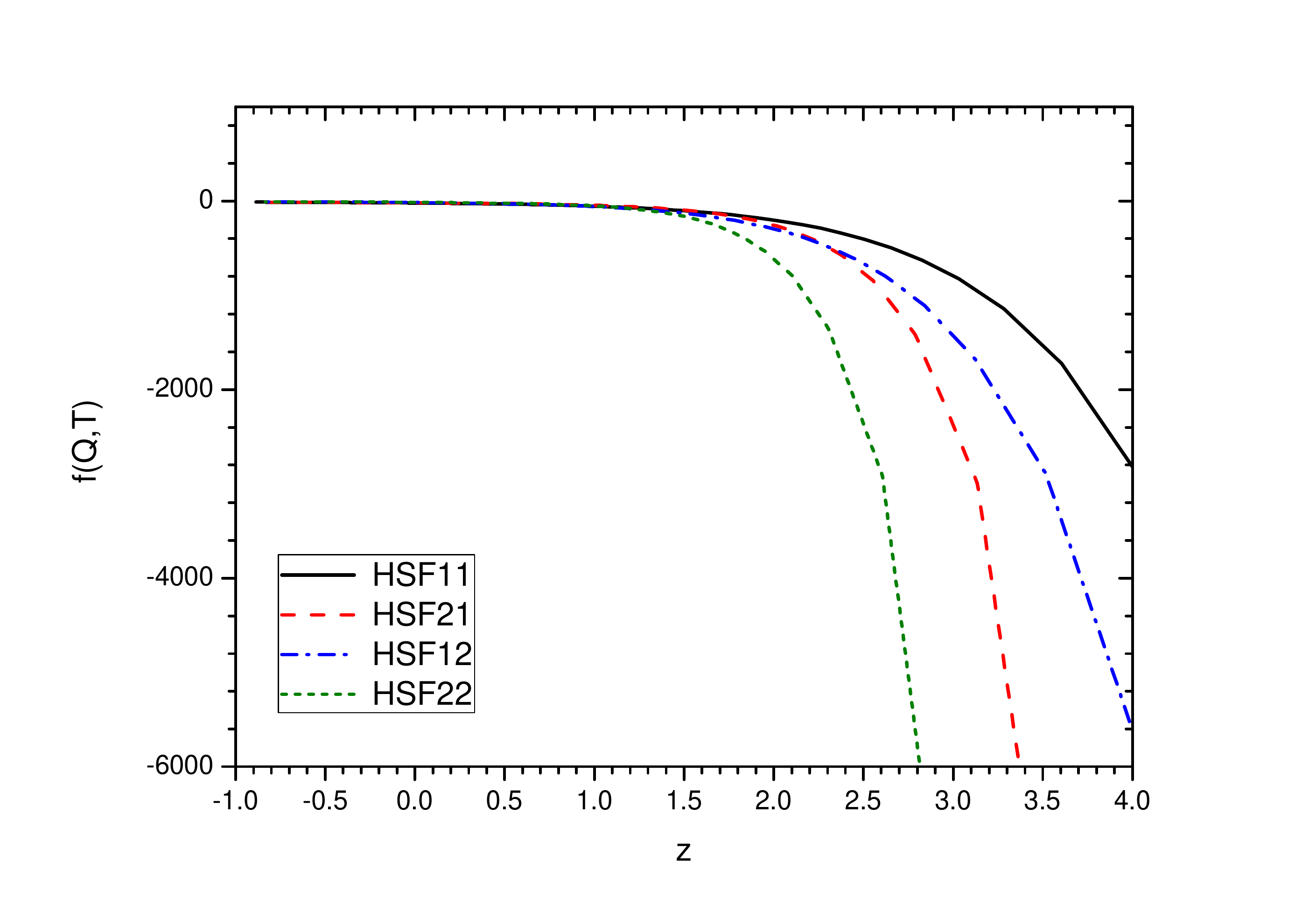}
\caption{The evolutionary behaviour of the functional $f(Q,T)$.}
\label{Fig4}
\end{figure}

\subsection{Case II}
As a second case, we consider the functional $f(Q,T)=aQ^{m}+bT$. One should note that, for $m=1$, this model reduces to the case discussed in the previous subsection. For this functional, we have $F=amQ^{m-1}$, $b=8\pi\kappa$, $\chi=amQ^{\left(m-1\right)}H$. Also, we have $\dot{F}=2(m-1)F\frac{\dot{H}}{H}$ and $\dot{\chi}=F\dot{H}\left(2m-1\right)$.

For this choice of the functional, the pressure and energy density becomes
\begin{eqnarray}
p &=& \frac{-\left(1-2m\right)aQ^m+2\dot{\chi}\left[2+\kappa-\kappa\kappa_1\right]}{4\pi\left[\left(2+\kappa\right)\left(2+3\kappa\right)-3\kappa^2\right]},\\
\rho &=& \frac{(1-2m)aQ^m+2\dot{\chi}\left[3\kappa-\left(2+3\kappa\right)\kappa_1\right]}{4\pi\left[\left(2+\kappa\right)\left(2+3\kappa\right)-3\kappa^2\right]}.
\end{eqnarray}

Consequently, the EoS parameter may be obtained from the above expressions of the pressure and energy density as
\begin{equation}
\omega=\frac{-\left(1-2m\right)aQ^m+2\dot{\chi}\left[2+\kappa-\kappa\kappa_1\right]}{(1-2m)aQ^m+2\dot{\chi}\left[3\kappa-\left(2+3\kappa\right)\kappa_1\right]}.
\end{equation}

One may note that, the above equations (20), (21) and (22) reduce to the corresponding equations of the previous case for $m=1$. For the HSF models, we may substitute in these equations, $Q=6\left(\alpha+\frac{\beta}{t}\right)^2$ and $\dot{\chi}=-am(2m-1)Q^{(m-1)}\frac{\beta}{t^2}$ to obtain the respective expressions for the pressure, energy density and the EoS parameter. Regarding the model parameters, we will consider the same parameter space for the HSF models. For the $f(Q,T)$ gravity model parameters, we consider $a=-4.4$, $b=0.1$ and some arbitrary values for $m$. We have checked that, for $m>0.5$, the energy density becomes positive and we get viable models. In view of this, we calculate the pressure, energy density and EoS parameter for five different values of $m$ namely $m=0.6,0.8, 1.1,1.5$ and $2$. It is to be noted that, for $m=1$, the model reduces to the previous model in subsection III-A. In all cases of $m$, we obtain similar behaviour. In view of this, we show the graphical representation of the energy density and EoS parameter for $m=0.6$ only. In FIG.5, the energy density is shown. The energy density remains in the positive domain throughout the cosmic evolution. For all the models, the energy density decreases smoothly from higher positive values to vanishingly small values at late times. In FIG. 6, the dynamical behaviour of the EoS parameter for the HSF models calculated within the framework of $f(Q,T)$ gravity is shown for a representative value of $m=0.6$. For all the models, the EoS parameter decreases from a positive value at an initial epoch to negative values at late epochs. It may be observed from the figure that, the HSF models coincide with the concordant $\Lambda$CDM value $\omega=-1$ at late times of cosmic evolution which is inconformity with recent observations. In order to assess the effect of the parameter $m$ on the cosmic dynamics, we have calculated the value of the EoS parameter at the present epoch, $\omega_0$, for different values of $m$. The calculated values are given in Table-I. One may note that, with an increase in the value of $m$, the EoS parameter in the present epoch increases.
\begin{table}
\caption{EoS parameter at the present epoch as predicted by the HSF models within $f(Q,T)$ gravity theory.}
\centering
\begin{tabular}{c|c|c|c|c}
\hline
\hline
Models & HSF11  & HSF21& HSF12 &HSF22\\
\hline
$m$		& $\omega_0$ &$\omega_0$ &$\omega_0$ &$\omega_0$ \\
\hline
0.6     & -0.847   & -0.889 &-0.798& -0.841\\
0.8     &-0.796   &-0.852&-0.730& -0.788\\
1.0		&-0.745&-0.815&-0.663&-0.736\\
1.1     &-0.719   &-0.797&-0.629&-0.709\\
1.5     &-0.617&-0.723&-0.494&-0.603\\
2	&-0.489&-0.631&-0.325&-0.471\\
\hline
\end{tabular}
\end{table}

\begin{figure}[!htp]
\centering
\includegraphics[scale=0.40]{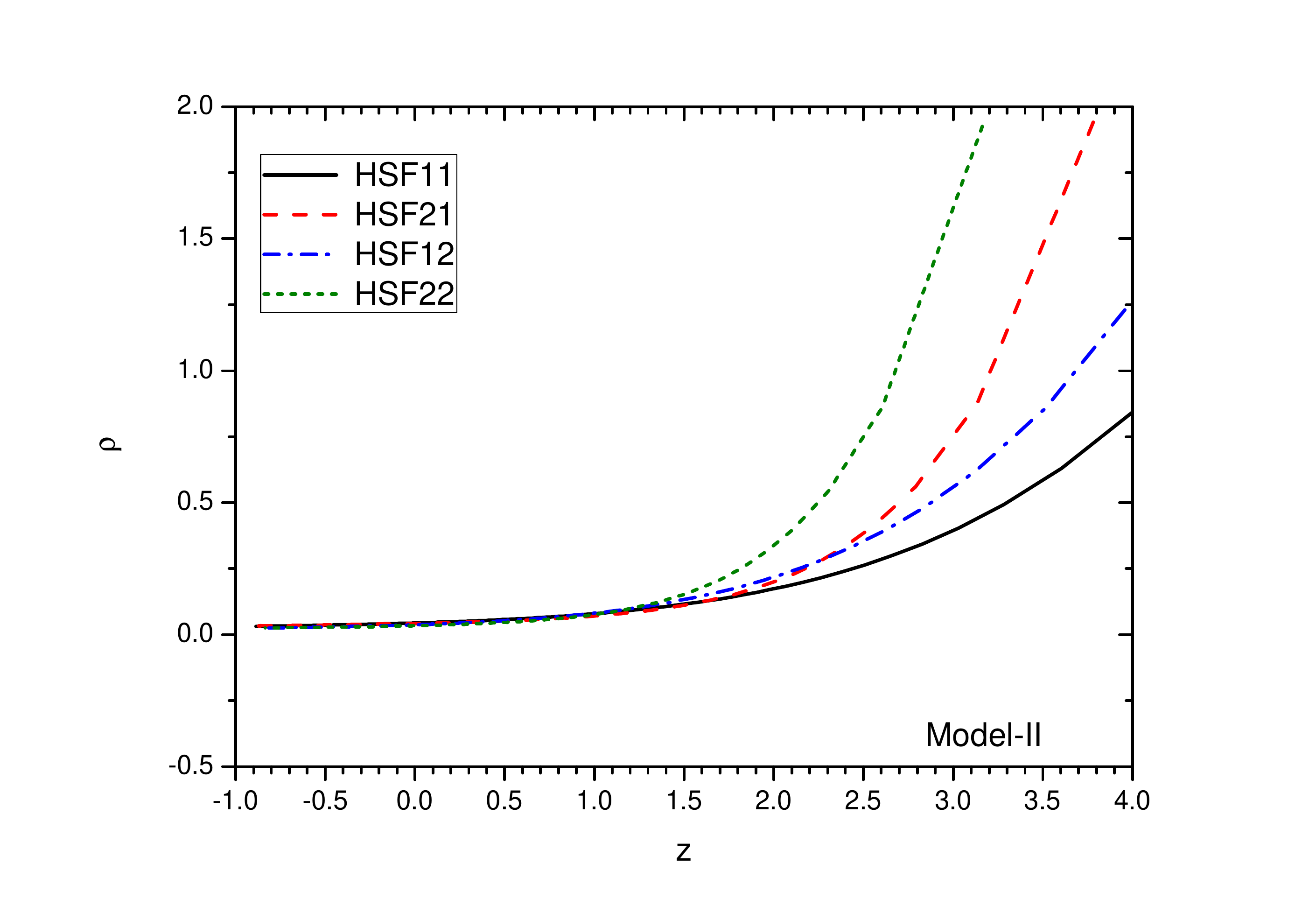}
\caption{ The evolutionary aspect of the energy density for the HSF models. Here we have used the parameter space $a=-4.4, b=0.01, m=0.6$.}
\label{Fig5}
\end{figure}

\begin{figure}[!htp]
\centering
\includegraphics[scale=0.40]{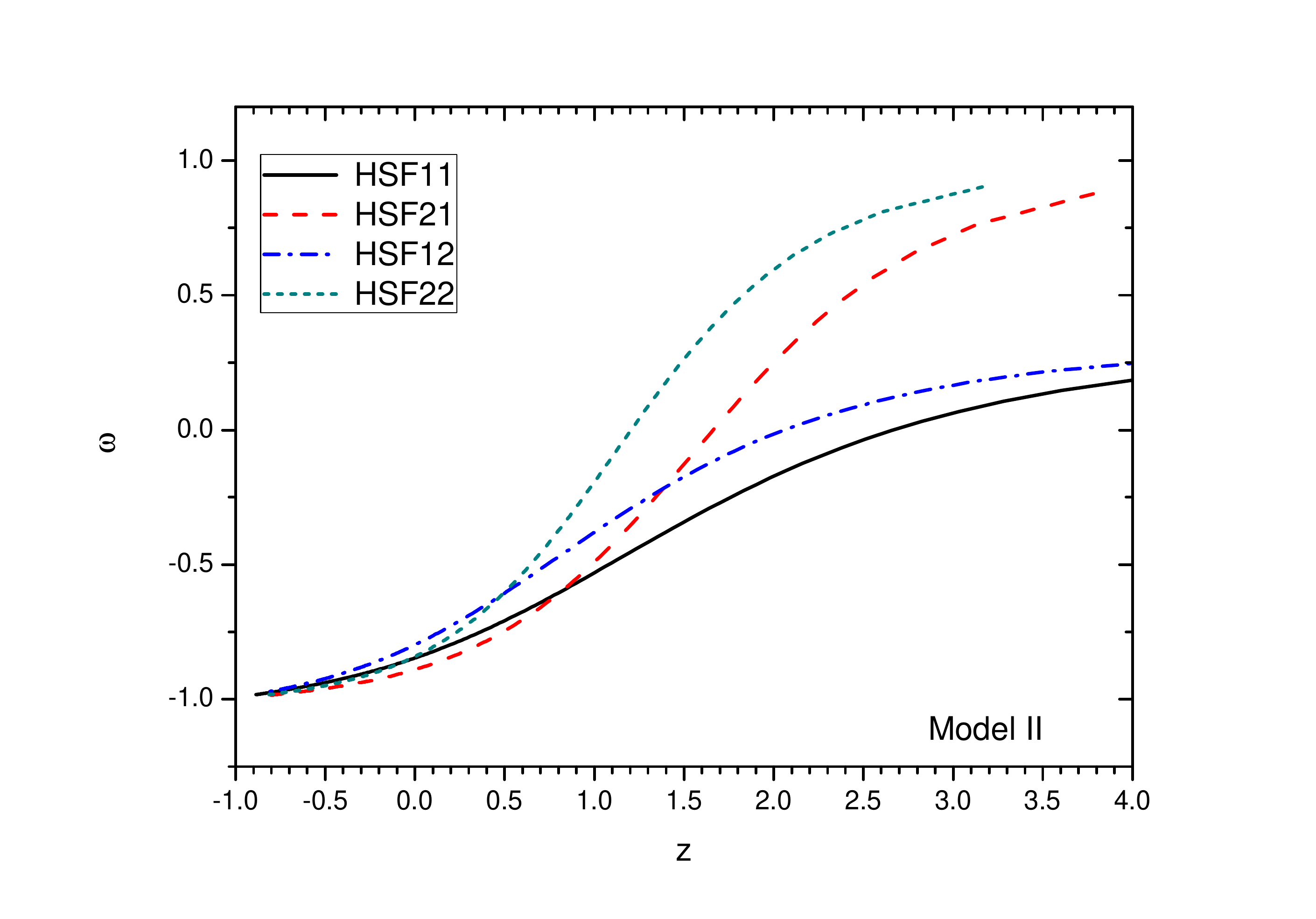}
\caption{The equation of state parameter in $f(Q,T)$ gravity. The parameter space used for the figure is $\left(a=-4.4, b=0.01, m=0.6\right)$.}
\label{Fig6}
\end{figure}

\begin{figure}[!htp]
\centering
\includegraphics[scale=0.40]{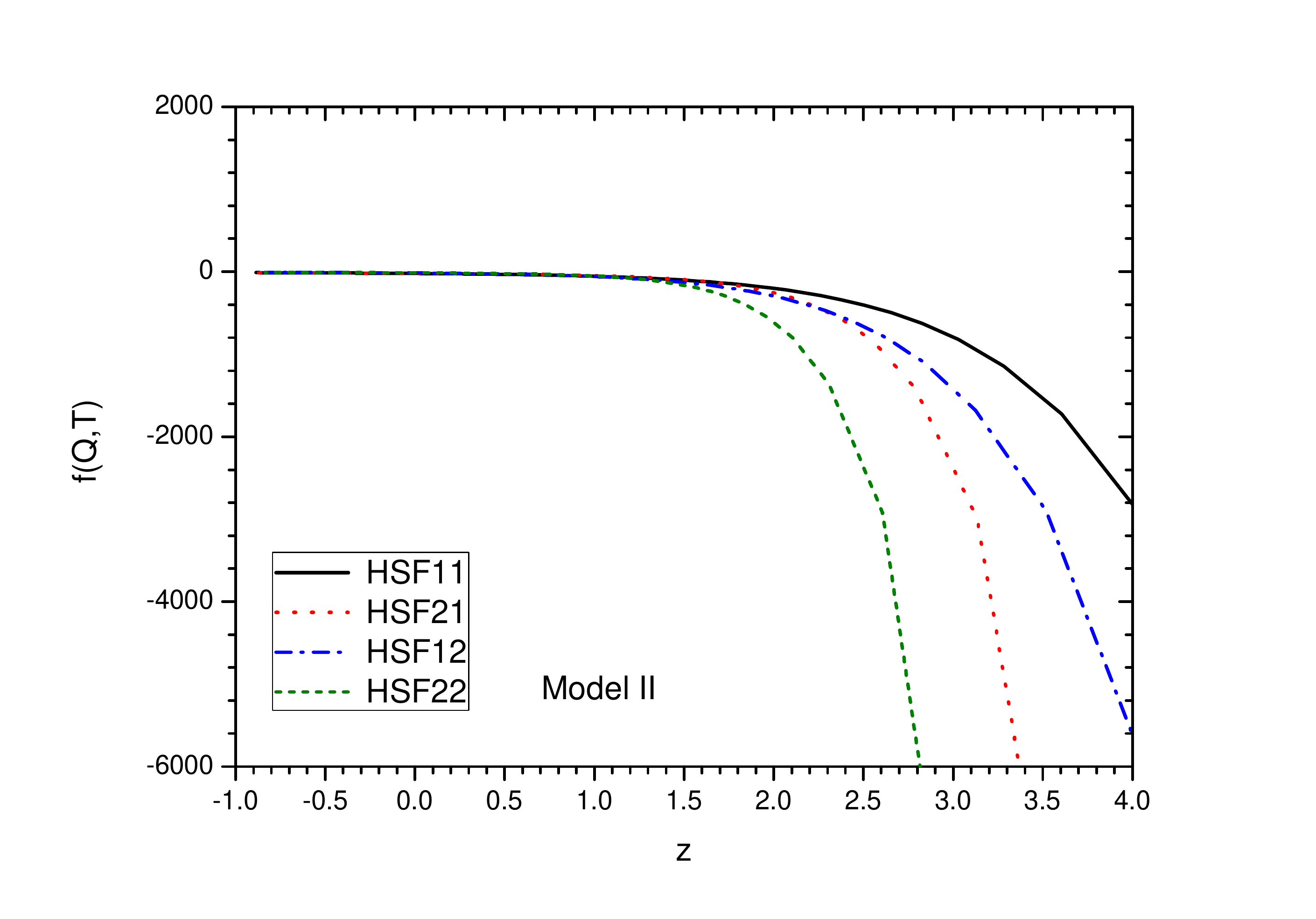}
\caption{The evolutionary behaviour of the functional $f(Q,T)$.}
\label{Fig7}
\end{figure}

In FIG.7, the functional $f(Q,T)=aQ^m+bT$ is shown as a function of redshift. For all the models, the functional assumes  negative values for the chosen parameter space. For any given HSF model, the functional is found to increase from lower negative values to small values at late times. Eventhough at an initial epoch, the values of the functional $f(Q,T)$ differ for the four different HSF models, at late epoch, all the HSF models predict almost same values for $f(Q,T)$. 
\section{Results and Conclusion}

In the present work, we have investigated some aspects of a modified  symmetric teleparallel gravity where a non-minimal coupling of the non-metricity $Q$ to the trace of the energy momentum tensor is assumed. Obviously, the $f(Q,T)$ leads to a non-conservation of energy-momentum tensor which may be responsible to achieve a late time cosmic acceleration in the models. The physical significance of the present theory is that, the appearance of an extra force term shoulders the burden of acceleration. This is essential in getting accelerating cosmological models without invoking any dark energy components to the matter field. It should be noted here that, the addition of some dark energy components in the matter field by hand may lead to ghost fields with negative kinetic energy terms in the Lagrangian. Two different functional forms of  $f(Q,T)$ are considered to model the Universe. Besides, we have used some recent constructed hybrid scale factors to investigate the cosmic dynamics. The HSF models provide a signature flipping behaviour of the deceleration parameter with early deceleration and late time acceleration.  Also, the HSF models predict reasonable values of the deceleration parameter at the present epoch as well as the values of transit redshift. For a given HSF model, we have derived the pressure, energy density and EoS parameter in most general form and studied their evolutionary behaviour. The model parameters are chosen suitably so as to provide a physically acceptable energy density. It is found from our investigations that, the model parameter $a$ does not affect the evolutionary aspect of the EoS parameter. However, the parameter $b$ affects marginally the behaviour of $\omega$. Another parameter $m$ that appears in the second $f(Q,T)$ model has a substantial effect on the EoS parameter in the sense that, with an increase in $m$, the value of $\omega$ at the present epoch increases.

As a final remark, we may say that, the present model is suitable extension of the symmetric teleparallel gravity and provide models that overlap with the concordant $\Lambda$CDM model at late times of cosmic evolution.  In view of this, these models may be useful in the context of the search for geometrical alternatives to dark energy.
\section*{Acknowledgement}
LP acknowledges Department of Science and Technology (DST), Govt. of India, New Delhi for awarding INSPIRE fellowship (File No. DST/INSPIRE Fellowship/2019/IF190600) to carry out the research work. We are thankful to the honourable referees for their constructive comments and suggestions for the improvement of the paper.


\begin{thebibliography}{99}
\section*{References}
\bibitem{Xu19} Y. Xu, G. Li,T. Harko, S. Liang, \textit{Eur. Phys. J. C}, \textbf{79}, 708 (2019).

\bibitem{Moller61} C. Moller, \textit{Mat. Fys. Skr. Dan. Vid. Selsk.}, \textbf{1}, 10 (1961).

\bibitem{Pellegrini63} C. Pellegrini, J. Plebanski, \textit{Mat. Fys. Skr. Dan. Vid. Selsk.}, \textbf{2}, 4 (1963).

\bibitem{Havashi79} K. Hayashi, T. Shirafuji, \textit{Phys. Rev. D}, \textbf{19}, 3524 (1979).

\bibitem{Aldrovandi13} R. Aldrovandi, J. G. Pereira, \textit{Teleparallel Gravity, Fundamental Theories of Physics} 173, Springer, Heidelberg, 2013.

\bibitem{Ferraro07} R. Ferraro, F. Fiorini, \textit{Phys. Rev. D}, \textbf{75}, 084031 (2007).

\bibitem{Linder10} E. V. Linder, \textit{Phys. Rev. D}, \textbf{81}, 127301 (2010).

\bibitem{Harko14} T.Harko, F. S. N. Lobo, G. Otalora, E. N. Saridakis, \textit{JCAP}, \textbf{12}, 021 (2014).

\bibitem{Capozziello17} S. Capozziello, G. Lambiase, E. N. Saridakis, \textit{Eur. Phys. J. C}, \textbf{77}, 576, (2017).

\bibitem{Otalora17} G. Otalora, M. J. Reboucas, \textit{Eur. Phys. J C}, \textbf{77}, 799 (2017).

\bibitem{Golovnev18} A. Golovnev, T. Koivisto, \textit{JCAP}, \textbf{11}, 012 (2018).

\bibitem{Fontanini19} M. Fontanini, E. Huguet, M. Le Delliou, \textit{Phys. Rev. D}, \textbf{99}, 064006 (2019).

\bibitem{Bose20} A. Bose, S. Chakraborty, \textit{Mod. Phys. Lett. A}, \textbf{35}, 2050296  (2020).

\bibitem{Golovnev20} A. Golovnev, M. Guzman, \textit{Phys. Lett. B}, \textbf{810}, 135806 (2020).

\bibitem{Nester99} J. M. Nester, H.-J. Yo, \textit{Chin. J. Phys.}, \textbf{37}, 113 (1999). 

\bibitem{Jimenez18} J. B. Jimenez, L. Heisenberg, T. Koivisto, \textit{Phys. Rev. D}, \textbf{98}, 044048 (2018).

\bibitem{Soudi18} I. Soudi et al., \textit{Phys. Rev. D}, \textbf{100}, 044008 (2019).

\bibitem{Lazkoz19} R. Lazkoz et al., \textit{Phys. Rev. D}, \textbf{100}, 104027 (2019).

\bibitem{Lu19} J. Lu, X. Zhao, G. Chee, \textit{Eur. Phys. J. C}, \textbf{79}, 530 (2019).

\bibitem{Bajardi20} F. Bajardi, D. Vernieri, S. Capozziello, \textit{Eur. Phys. J. Plus}, \textbf{135}, 912 (2020).

\bibitem{Frusciante21} N. Frusciante, \textit{Phys. Rev. D}, \textit{103}, 044021 (2021).

\bibitem{Xu20} Y. Xu et al., \textit{Eur. Phys. J. C}, \textbf{80}, 449 (2020).

\bibitem{Zia21} R. Zia, D. C. Maurya, A. K. Shukla, \textit{Int. J. Geom. Methods Phys.}, \textbf{18}, 2150051 (2021).

\bibitem{Mishra15} B. Mishra and S. K. Tripathy, \textit{Mod. Phys. Lett. A} \textbf{36}, 1550175 (2015).

\bibitem{Mishra18a} B. Mishra, S. K. Tripathy and P. P. Ray, \textit{Astrophys. Space Sci.} \textbf{363}, 86 (2018).

\bibitem{Mishra18} B. Mishra, S.K. Tripathy, S. Tarai,\textit{Mod. Phys. Lett. A}, \textbf{33}, 1850052 (2018).

\bibitem{Mishra19} B. Mishra, P. P. Ray and R. Myrzakulov, \textit{Eur. Phys. J. C} \textbf{79}, 34 (2019).

\bibitem{Mishra19a} B. Mishra, P. P. Ray, S.K. Tripathy and K. Bamba, \textit{Mod. Phys. Lett. A} \textbf{27}, 1950217 (2019)

\bibitem{Ray19} P. P. Ray, B. Mishra and S.K. Tripathy, \textit{Int. J. Mod. Phys. D} \textbf{28}, 1950093 (2019).

\bibitem{Tripathy2020} S. K. Tripathy, S. K. Pradhan, Z. Naik, D. Behera and B. Mishra, \textit{Phys. of Dark Univ.} \textbf{30}, 100722 (2020).

\bibitem{SKT2020} S. K. Tripathy, S. K. Pradhan, P. Parida, D. Behera, R. K. Khuntia and B. Mishra, \textit{Phys. Scr.}, \textbf{95}, 115001 (2020).

\bibitem{Mishra21} B. Mishra, S.K. Tripathy and S. Tarai, J. Astrophys. Astron., \textbf{42}, 2 (2021).

\bibitem{SKT2021} S. K. Tripathy, B. Mishra, M. Khlopov and S. Ray, \textit{Int. J. Mod. Phys. D}:DOI: 10.1142/S0218271821400058 (2021).

\bibitem{Reiss2016}  A. G. Reiss et al., \textit{Astrophys. J.} \textbf{826(1)}, 56 (2016). arXiv:1604.01424

\bibitem{Reiss2018}  A. G. Riess et al., \textit{Astrophys. J.} \textbf{861(2)}, 126 (2018). arXiv:1804.10655

\bibitem{Aghanim18} N. Aghanim et al, \textit{Astron. Astrophys.} \textbf{641}, A6 (2020). Planck Collaboration.)

\bibitem{Reid2019} M. J. Reid, D. W. Pesce and A. G. Riess, \textit{Astrophys. J.} \textbf{886(2)}, L27 (2019). arXiv:1908.05625

\bibitem {Busca13}N. G. Busca, \textit{Astron. Astrophys}, \textbf{552}, A96(2013).

\bibitem {Farooq13}O. Farooq and B. Ratra, \textit{Astrophys J. Lett.}, \textbf{766}, L7(2013).

\bibitem {Moresco16}M. Moresco \textit{et al}., \textit{J. Cosmol. Astropart. Phys.}, \textbf{2016}, 1605, 014(2016).

\bibitem{Amanullah10} R. Amanullah et al., \textit{Astrophys. J.}, \textbf{716}, 712 (2010).

\bibitem{Hinshaw13} G. Hinshaw et al., \textit{Astrophys. J. Suppl. Ser.}, \textbf{208}, 19 (2013).

\end{thebibliography}
\end{document}